\documentclass[letterpaper, 10 pt, conference]{ieeeconf}
\IEEEoverridecommandlockouts  
\usepackage{graphicx} 
\usepackage{cite}
\usepackage{caption}
\usepackage{amsmath}
\usepackage{amsfonts}
\usepackage{dsfont}
\let\labelindent\relax
\usepackage{enumitem,kantlipsum}
\usepackage{subcaption}
\usepackage{bm}
\usepackage{amsmath,amssymb}
\usepackage{threeparttable}
\usepackage{xcolor}

\newtheorem{remark}{Remark}

\date{} 

\title{\LARGE \bf Exact-Time Safety Recovery using Time-Varying Control Barrier Functions with Optimal Barrier Tracking}

\author{Yingqing Chen, Christos G. Cassandras, Wei Xiao and Anni Li
\thanks{Y. Chen and C. G. Cassandras are
with the Division of Systems Engineering and Center for Information and
Systems Engineering, Boston University, Brookline, MA 02446
\tt\small\{yqchenn;cgc\}@bu.edu.}
\thanks{Wei Xiao is with the robotics engineering department at WPI and with CSAIL, MIT. \texttt{{\small weixy@mit.edu}}}
\thanks{Anni Li is with the ECE department at UNC-Charlotte. \texttt{{\small ali20@charlotte.edu}} }
}

\begin{document}
\maketitle
\thispagestyle{empty}
\pagestyle{empty}

\begin{abstract} 
This paper is motivated by controllers developed for autonomous vehicles which occasionally lead into conditions where safety is no longer guaranteed. We develop an exact-time safety recovery framework for any control-affine nonlinear system when its state is outside a safe region using time-varying Control Barrier Functions (CBFs) with optimal barrier tracking. Unlike conventional formulations that provide only conservative upper bounds on recovery time convergence, the proposed approach guarantees recovery to the safe set at a prescribed time. The key mechanism is an active barrier tracking condition that forces the barrier function to follow exactly a designer-specified recovery trajectory. This transforms safety recovery into a trajectory design problem. The recovery trajectory is parameterized and optimized to achieve optimal performance while preserving feasibility under input constraints,  avoiding the aggressive corrective actions typically induced by conventional finite-time formulations. The safety recovery framework is applied to the roundabout traffic coordination problem for Connected and Automated Vehicles (CAVs), where any initially violated safe merging constraint is replaced by an exact-time recovery barrier constraint to ensure safety guarantee restoration before subsequent CAV conflict points are reached. Simulation results demonstrate improved feasibility and performance.
 
\end{abstract}



\section{Introduction}
Safety-constrained optimization problems commonly arise in cyber-physical systems
with autonomous agents.
This is particularly true in transportation systems involving Connected and Automated Vehicles (CAVs) with real-time V2X communication \cite{liu2025serverless}:
CAV controllers must satisfy safety constraints while simultaneously seeking to optimize performance objectives under input limits, uncertain arrivals, and strong vehicle interactions \cite{rios2017survey,rios2017merging,campi2023roundabouts}.

Control Barrier Functions (CBFs) are widely used to capture safety constraints in control design. Their theoretical basis dates back to Nagumo's theorem \cite{nagumo1942lage}: if a safety function is nonnegative at initialization and does not decrease on the boundary, the corresponding safe set is forward invariant. Standard CBF formulations implement this condition through class-$\mathcal{K}$ functions \cite{Ames2017TAC, glotfelter2017nonsmooth} keeping the barrier value away from the boundary while still preserving forward invariance. For higher-relative-degree constraints, High-Order CBF (HOCBF) constructions extend the same idea to more general nonlinear systems \cite{nguyen2016exponential,xiao2022hocbf}.
CBF and HOCBF methods have been widely adopted in optimization-based safety-critical control for CAV conflict areas, including merging roadways \cite{milanes2010automated}, lane changing \cite{li2025robust}, and roundabout coordination \cite{xu2022decentralized}. 
However, most existing formulations assume safe initialization, i.e., the system state starts inside the safe set.

When initial conditions are unsafe, existing CBF-based methods follow several  different approaches. Backup-policy approaches, including backup CBF and multi-backup designs, preserve safety by requiring at least one predesigned fallback controller to remain feasible \cite{chen2021backup,janwani2024learning}. Robustness-oriented variants handle uncertainty by tightening the CBF inequality, for example through worst-case margins or stricter admissible-control conditions, so that the certified safe set remains forward invariant under sensing and model errors \cite{cosner2021measurement,garg2021robust}. Prescribed-time recovery approaches introduce explicit time dependence into barrier dynamics, such as prescribed-time gains or time-varying shaping terms, to drive the barrier value toward nonnegativity within a selected horizon \cite{zhang2024prescribed,huang2024learning}. Similarly, related 
Control Lyapunov Barrier Function (CLBF)
and High Order CLBF (HOCLBF)
constructions use convergence-shaping terms to pull unsafe states toward the boundary \cite{xiao2021clbf,chen2025optimal}. Although this line of work improves recoverability, many methods still rely on fixed convergence structures or conservative upper bounds on recovery time, which can lead to overly conservative transients under input constraints, or even infeasibility.

To overcome these limitations, this paper develops an exact-time safety recovery framework for control-affine systems with unsafe initial conditions. The core idea is a time-varying recovery CBF that enforces tracking of a prescribed barrier trajectory. This mechanism converts safety recovery from a rate-bound convergence condition into a trajectory design problem in barrier space. By parameterizing the recovery trajectory and optimizing its shape under input constraints, the framework provides direct control of recovery transients while guaranteeing convergence to the safe set at an exact prescribed time
(when this is feasible under existing control constraints).

Our 
contributions are fourfold.
First, an Exact-Time Recovery CBF (ExT-CBF) condition is established to guarantee recovery at a prescribed time rather than within a conservative upper bound. Second, an exact barrier-tracking property under active enforcement is derived, which provides a direct mechanism for transient shaping. 
Third, the recovery trajectory is parameterized and optimized under input constraints to support practical trade-offs between feasibility and control performance. 
Fourth, the framework is applied to the problem of coordinating CAVs in a roundabout. 
As observed in 
\cite{chen2025optimal},
when CAVs in a roundabout transition from one segment (control zone) to the next, their initial conditions in the next segment frequently fail to guarantee safe merging at the next conflict point. We show how to restore exactly this safe merging constraint with simulation results showing improved feasibility and performance compared with conventional finite-time recovery formulations.

The remainder of the paper is organized as follows. Section II introduces preliminaries and formulates the exact-time safety recovery problem. Section III presents the ExT-CBF construction and the exact barrier-tracking property. Section IV develops the optimal barrier-tracking formulation. Section V applies the method to CAV roundabout control. Section VI reports simulation results.

\section{Preliminaries and Problem Formulation}

\subsection{Barrier-Based Safety Control}

Consider the control-affine system
\begin{equation}
\dot{\bm{x}} = f(\bm{x}) + g(\bm{x})\bm{u},
\label{eq:system}
\end{equation}
where $\bm{x} \in \mathbb{R}^n$, $\bm{u} \in U \subset \mathbb{R}^p$, 
$f : \mathbb{R}^n \rightarrow \mathbb{R}^n$ and 
$g : \mathbb{R}^n \rightarrow \mathbb{R}^{n \times p}$ are locally Lipschitz, and $U$ is compact and convex. 

Let $b:\mathbb{R}^n \rightarrow \mathbb{R}$ be a continuously differentiable function defining the safe set
\begin{equation}
C := \{ \bm{x} \in \mathbb{R}^n \mid b(\bm{x}) \ge 0 \}.
\label{eq:safeset}
\end{equation}

\textbf{Control Barrier Functions.}
Following \cite{Ames2017TAC, xiao2023safe}, $b(\bm{x})$ is a Control Barrier Function (CBF) for \eqref{eq:system} if there exists an extended class-$\mathcal{K}$ function $\alpha(\cdot)$ such that
\begin{equation}
\sup_{\bm{u} \in U}
\left[
L_f b(\bm{x}) + L_g b(\bm{x}) \bm{u}
+ \alpha(b(\bm{x}))
\right] \ge 0,
\label{eq:cbf}
\end{equation}
where $L_f b(x)$ and $L_g b(x)$ denote the Lie derivatives of $b$ along $f$ and $g$, respectively.
Any locally Lipschitz controller satisfying \eqref{eq:cbf} renders $C$ forward invariant provided $b(\bm{x}(0)) \ge 0$.

\textbf{Relative degree.} 
The relative degree of a function $b(\bm x)$ (or a constraint $b(\bm x)\geq 0$) w.r.t. (\ref{eq:system}) is defined \cite{xiao2023safe}
as the minimum number of times we need to differentiate $b(\bm x)$ along system (\ref{eq:system}) until any control component of $\bm u$ explicitly shows up in the corresponding derivative.

When $b(\bm{x})$ has relative degree $m>1$, High Order CBFs (HOCBFs) extend this framework by recursively constructing auxiliary functions to accommodate arbitrary relative degree constraints \cite{xiao2022hocbf}. These constructions guarantee forward invariance of appropriately defined constraint sets when the initial condition satisfies the safety requirement. 

\subsection{Recovery from Safety Violations.}
Classical CBF and HOCBF formulations assume
$b(\bm{x}(0)) \ge 0$.
However, in many safety-critical applications such as vehicle coordination and collision avoidance, the system may be initialized outside the safe set
or enter an unsafe region that was not previously detected, 
 i.e., 
\begin{equation}\label{eq:unsafe_init}
    b(\bm{x}(0)) < 0
\end{equation}

One approach to address this situation is to modify the class-$\mathcal{K}$ function in the CBF condition to induce convergence toward the safe set. In particular, Control Lyapunov Barrier Functions (CLBFs) \cite{xiao2021clbf} incorporate stability and safety requirements by selecting nonlinear class-$\mathcal{K}$ functions of the form
\begin{equation}
\alpha(b(\bm{x})) 
= p(b(\bm{x}))^q,
\label{eq:clbf_class_k}
\end{equation}
where $p>0$ and $0<q<1$. Under this construction, the barrier dynamics enforce finite-time convergence of $b(\bm{x}(t))$ to zero, yielding recovery of the safety constraint within a computable upper bound.

While \eqref{eq:clbf_class_k} guarantees finite-time convergence, the recovery time is characterized by an upper bound determined by the decay parameters and initial condition, rather than a prescribed exact time. Moreover, near the safe-set boundary, the corresponding derivative can become unbounded, which may induce chattering behavior \cite{xiao2021clbf}. We therefore consider an exact time safety recovery problem formulation as follows.

\textbf{Exact-Time Safety Recovery Problem.}
Consider system \eqref{eq:system} over a finite horizon $[t_0,\, t_f]$, where $t_f>t_0$ is a given terminal time. The performance objective is
\begin{equation}
\min_{\bm{u}(\cdot)} 
\int_{t_0}^{t_f} 
\ell(\bm{x}(t),\bm{u}(t))\,dt,
\label{eq:ocp}
\end{equation}
subject to $\bm{u}(t)\in U$ for all $t\in[t_0,t_f]$.

Suppose the system is initialized outside the safe set at time $t_0$,
\begin{equation}
b(\bm{x}(t_0)) = b_0 < 0.
\end{equation}
Given a prescribed recovery time $t_1 \in (t_0, t_f]$, equivalently a recovery horizon $T_r := t_1 - t_0 > 0$, the objective is to design a control input $\bm{u}(t)\in U$ such that
\begin{align}
b(\bm{x}(t)) &< 0, \quad \forall t \in [t_0,\,t_1), \\
b(\bm{x}(t_1)) &= 0, \\
b(\bm{x}(t)) &\ge 0, 
\quad \forall t \in (t_1,\,t_f],
\label{eq:exact_recovery}
\end{align}
while satisfying the system dynamics and optimizing the objective \eqref{eq:ocp}.

To enforce the exact-time recovery condition above, we introduce a time-varying control barrier function construction. The key idea is to embed the prescribed recovery time into a time-dependent barrier condition such that the evolution of $b(\bm{x}(t))$ is explicitly regulated over $[t_0, t_1]$. 

\section{Exact-Time Recovery via Time-Varying CBF}\label{sec: Exact-Time Recovery via Time-Varying CBF}

To solve the Exact-Time Safety Recovery Problem formulated above, we embed the prescribed recovery time directly into a time-varying barrier function. The central idea is to regulate the evolution of the barrier value itself over $[t_0,t_1]$, so that \emph{recovery is achieved exactly at the prescribed time} rather than within a conservative upper bound.

Let $\gamma:[t_0,t_1]\to\mathbb{R}$ be a continuously differentiable \emph{recovery function} satisfying
\begin{equation}
\gamma(t_0)=b_0, 
\qquad 
\gamma(t_1)=0, 
\qquad 
\dot{\gamma}(t)>0, \ \forall t\in[t_0,t_1).
\label{eq:gamma}
\end{equation}
The function $\gamma(t)$ prescribes the desired recovery from the initial violation to the boundary of the safe set at time $t_1$. 

Define the shifted barrier function
\begin{equation}
s(\bm{x},t) := b(\bm{x}(t)) - \gamma(t).
\label{eq:shifted_s}
\end{equation}
If $s(\bm{x}(t),t)\ge 0$ for all $t\in[t_0,t_1]$, then
$b(\bm{x}(t))\ge \gamma(t)$, and in particular
$b(\bm{x}(t_1))\ge 0$, ensuring recovery at the prescribed time. Therefore, it suffices to render the time-varying set 
$\hat C(t):=\{\bm{x}\mid s(\bm{x},t)\ge 0\}$ forward invariant on $[t_0,t_1]$.
Differentiating \eqref{eq:shifted_s} along \eqref{eq:system} yields
\begin{equation}
\dot s(\bm{x},t)
=
L_f b(\bm{x}(t)) + L_g b(\bm{x}(t))\bm{u}
- \dot{\gamma}(t).
\label{eq:s_dot_ext}
\end{equation}

Replacing the barrier function in \eqref{eq:cbf} with the shifted barrier function \eqref{eq:shifted_s}, we impose the time-varying CBF condition that for all $t\in[t_0,t_1]$:
\begin{equation}
L_f b(\bm{x}) + L_g b(\bm{x})\bm{u}
- \dot{\gamma}(t)
+ \alpha\!\left(b(\bm{x})-\gamma(t)\right)
\ge 0,
\label{eq:ext_cbf}
\end{equation}

\textbf{Definition 1 [Exact-Time Recovery CBF (ExT-CBF)].}
A 
continuously differentiable function $b : \mathbb{R}^n \to \mathbb{R}$ is a \emph{Recovery CBF} for system \eqref{eq:system} on $[t_0,t_1]$ with respect to recovery function $\gamma(t)$ if initial condition \eqref{eq:unsafe_init} holds and \eqref{eq:gamma} is satisfied, and there exists a class-$\mathcal{K}$ function $\alpha$ such that 
\begin{equation}
\sup_{\bm{u}\in U}
\Big[
L_f b(\bm{x}) + L_g b(\bm{x})\bm{u}
- \dot{\gamma}(t)
+ \alpha\!\left(b(\bm{x})-\gamma(t)\right)
\Big]
\ge 0 
\label{eq:rbf}
\end{equation}
for all $t\in[t_0,t_1]$.

If, in addition, there exists $\bm{u}(t)\in U$ such that
\begin{equation}
L_f b(\bm{x}) + L_g b(\bm{x})\bm{u}
- \dot{\gamma}(t)
+ \alpha\!\left(b(\bm{x})-\gamma(t)\right)
= 0
\label{eq:ext_cbf_equal}
\end{equation}
holds for all $t\in[t_0,t_1]$ along the resulting trajectory $\bm{x}(t)$, then $b(\bm{x}(t))$ is called an
\emph{Exact-Time Recovery CBF (ExT-CBF)} on $[t_0,t_1]$ under the controller
$\bm{u}(t)$.


\textbf{Lemma 1 [Exact Barrier Tracking].}
Let $b(\bm{x}(t))$ be an ExT-CBF under corresponding controller on $[t_0,t_1]$ with recovery function $\gamma(\cdot)$ satisfying \eqref{eq:gamma}. Then
\begin{equation}
b(\bm{x}(t))=\gamma(t),
\qquad \forall t\in[t_0,t_1].
\label{eq:exact_tracking}
\end{equation}

\begin{proof}
With tracking error $s(\bm{x}(t),t)$ defined in \eqref{eq:shifted_s},
Since $b(\cdot)$ is continuously differentiable and $\bm{x}(\cdot)$ is continuous, $s(\bm{x}(t),t)$ is differentiable almost everywhere and
\begin{equation}
    \dot{s}(\bm{x}(t),t) = \dot{b}(\bm{x}(t)) - \dot{\gamma}(t).
\end{equation}

Since the $b(\bm{x}(t))$ is ExT-CBF, it follows from Definition 1 that condition \eqref{eq:ext_cbf_equal} holds.
Using the chain rule and \eqref{eq:system}, 
\begin{equation}
  \dot{b}(x(t))
= \dot{\gamma}(t) - \alpha\big(b(x(t)) - \gamma(t)\big).  
\end{equation}

Substituting into the error dynamics yields
\begin{equation} \label{eqn: active dot_s}
    \dot{s}(\bm{x}(t),t)
= -\alpha (s(\bm{x}(t),t)).
\end{equation}

Because $\alpha(\cdot)$ is a class-$\mathcal{K}$ function, and  ${s}(\bm{x}(t),t)\geq 0 $ by the forward invariance property of \eqref{eq:ext_cbf}, 
we have $\dot{s}(\bm{x}(t),t) \leq 0$ for $t \in [t_0, t_1]$ by \eqref{eqn: active dot_s}.
Hence $s(\bm{x}(t),t)= 0$ is the unique solution of \eqref{eqn: active dot_s} with $s(t_0)=0$, which implies \eqref{eq:exact_tracking}
\end{proof}

The following theorem shows that 
recovery to the boundary of the safe set occurs exactly at the prescribed time $t_1$.

\textbf{Theorem 1 [Exact-Time Safety Recovery].}
Suppose $b(\bm{x}(t))$ is an ExT-CBF on $[t_0,t_1]$ with recovery function $\gamma(\cdot)$ satisfying \eqref{eq:gamma}. Then
\begin{equation}
b(\bm{x}(t_1)) = 0.
\end{equation}

\emph{Proof.}
By Lemma 1, $b(\bm{x}(t))=\gamma(t)$ for all $t\in[t_0,t_1]$. Evaluating at $t_1$ and using $\gamma(t_1)=0$ yields $b(\bm{x}(t_1))=0$. \hfill $\blacksquare$

\begin{remark}[Exact-time versus finite-time recovery] When the ExT-CBF constraint is rendered active, exact barrier tracking holds and recovery occurs precisely at the prescribed time $t_1$. 
In contrast, the more general recovery CBF (as in Definition 1)
without enforcing that it be active guarantees finite-time recovery, where $t_1$ serves only as an upper bound on the recovery time. 
\end{remark}

The preceding results show that once the recovery constraint is rendered active, the barrier value evolves exactly along the prescribed trajectory $\gamma(t)$ and reaches the safe boundary at the specified time. Recovery is therefore achieved by construction, rather than inferred from a conservative convergence bound. This mechanism removes the structural rigidity imposed by fixed polynomial decay laws such as \eqref{eq:clbf_class_k} and reframes safety recovery as a trajectory design problem in the barrier space. Consequently, by parameterizing $\gamma(t)$ and optimizing its shape under input constraints, one can explicitly regulate the recovery profile while balancing feasibility and performance objectives. This observation naturally motivates the development of an optimal barrier tracking formulation in the next section.

\section{Optimal Barrier Tracking Formulation}

Because the barrier trajectory follows $\gamma(t)$ during recovery, we formulate an optimal barrier tracking problem in which the recovery function is parameterized and designed to achieve a desired \emph{balance between feasibility and performance.}

\textbf{Parameterized Recovery Function.}
To enable systematic design, we parameterize the recovery function as $\gamma(t;\theta)$ with $\theta \in \mathbb{R}^r$, where $r$ denotes the number of free parameters in the recovery profile, subject to the same boundary conditions \eqref{eq:gamma}.
The parameter vector $\theta$ determines the recovery shape, while the boundary conditions enforce exact-time convergence.

A convenient representation consists of the normalized time 
$\tau := (t-t_0)/T_r \in [0,1]$, where $T_r=t_1-t_0$, and defines
\begin{equation}
\gamma(t;\theta)=b_0\,\phi(\tau;\theta),
\end{equation}
with $\phi(0;\theta)=1$ and $\phi(1;\theta)=0$. The kernel 
$\phi(\tau;\theta)$ determines the recovery profile. Typical choices include (a) quadratic forms such as $\phi(\tau)=1-a\tau-b\tau^2$, where higher-order polynomials provide additional curvature control; and (b) exponential kernels
$\phi(\tau)=\frac{e^{-k\tau}-e^{-k}}{1-e^{-k}}$,
where $k>0$ regulates front-loaded versus back-loaded recovery. More aggressive kernels accelerate early correction but may increase the control effort or compromise feasibility under input bounds, while smoother kernels distribute recovery more evenly. This parameterization provides a direct mechanism to balance recovery rate and feasibility.

\textbf{Parameterized Recovery Optimization.}
Since exact barrier tracking implies $b(\bm{x}(t))=\gamma(t;\theta)$, the parameter vector $\theta$ fully determines the safety recovery trajectory and indirectly determines the control input through the ExT-CBF equality condition. 
We therefore formulate \textbf{Problem 1}:
\begin{equation}
\begin{aligned}
\min_{\theta \in \mathbb{R}^r, \bm{u}(t)} \quad 
& \int_{t_0}^{t_1} J\big(\theta,t\big)\,dt \\
\text{s.t.}\quad 
& \gamma(t_0;\theta)=b_0, \quad \gamma(t_1;\theta)=0, \\
& \dot{\gamma}(t;\theta)>0, \\
& L_f b(\bm{x}(t)) + L_g b(\bm{x}(t))\bm{u}(t)
= \dot{\gamma}(t;\theta), \\
& \bm{u}(t)\in U,
\quad \eqref{eq:system}, \quad \forall t\in[t_0,t_1],
\end{aligned}
\label{eq:parameterized_recovery}
\end{equation}
where $J(\theta,t)$ captures performance objectives such as control effort, smoothness, terminal speed deviation, or weighted combinations.

Because the barrier dynamics are explicitly prescribed, optimizing $\theta$ directly determines the recovery trajectory while preserving exact-time convergence. Feasibility of the tracking equality is inherently enforced by the input constraint $u(t)\in U$, which implicitly restricts admissible recovery profiles.

\begin{remark}[ExT-CBF versus CLBF]
The proposed ExT-CBF generalizes CLBF-based recovery by replacing the fixed polynomial decay law 
\[
L_f b(x)+L_g b(x)u = p(-b(x))^{q}
\] 
with the tracking condition 
\[
L_f b(x)+L_g b(x)u = \dot{\gamma}(t).
\]
Hence, a CLBF can be viewed as a special case corresponding to a particular polynomial choice of $\gamma(t)$. 

In addition, unlike CLBF, whose fixed decay structure may conflict with input bounds and lead to conservativeness or infeasibility, ExT-CBF parameterizes $\gamma(t)$ and thus adapts the recovery profile to remain compatible with control constraints. Moreover, when enforced actively, it guarantees exact barrier tracking and achieves recovery at a designer-specified time, rather than within a conservative upper bound. 

Related recovery formulations in \cite{li2026finitetimeconvergentcontrolbarrier, gadginmath2026constrictingtubesprescribedtimesafe} address similar limitations of CLBF-type methods, but with different emphases. The framework in \cite{li2026finitetimeconvergentcontrolbarrier} guarantees convergence within an upper time bound and emphasizes feasibility conditions and parameter selection for constructing a recovery CBF under control bounds, whereas \cite{gadginmath2026constrictingtubesprescribedtimesafe} imposes prescribed-time recovery through a shrinking tube with a feasibility characterization. In contrast, our ExT-CBF framework treats recovery as a barrier-trajectory design problem, guaranteeing exact recovery at a prescribed time through active tracking while allowing the recovery profile itself to be parameterized and optimized under input constraints.
\end{remark}

Through parameter optimization, the recovery profile generated by the ExT-CBF controller can be shaped to satisfy application-specific performance and feasibility requirements. By decoupling recovery time specification from trajectory design, the framework offers a flexible and implementable mechanism for safety restoration under input constraints.

We next apply the ExT-CBF approach to the roundabout vehicle coordination problem, where merging conflict constraints may be violated due to stochastic vehicle arrivals or CZ transition.

\section{Optimal Control for Roundabout}\label{Sec:Optimal Control for Roundabout}
\begin{figure}
    \centering
    \includegraphics[width=0.7\columnwidth]{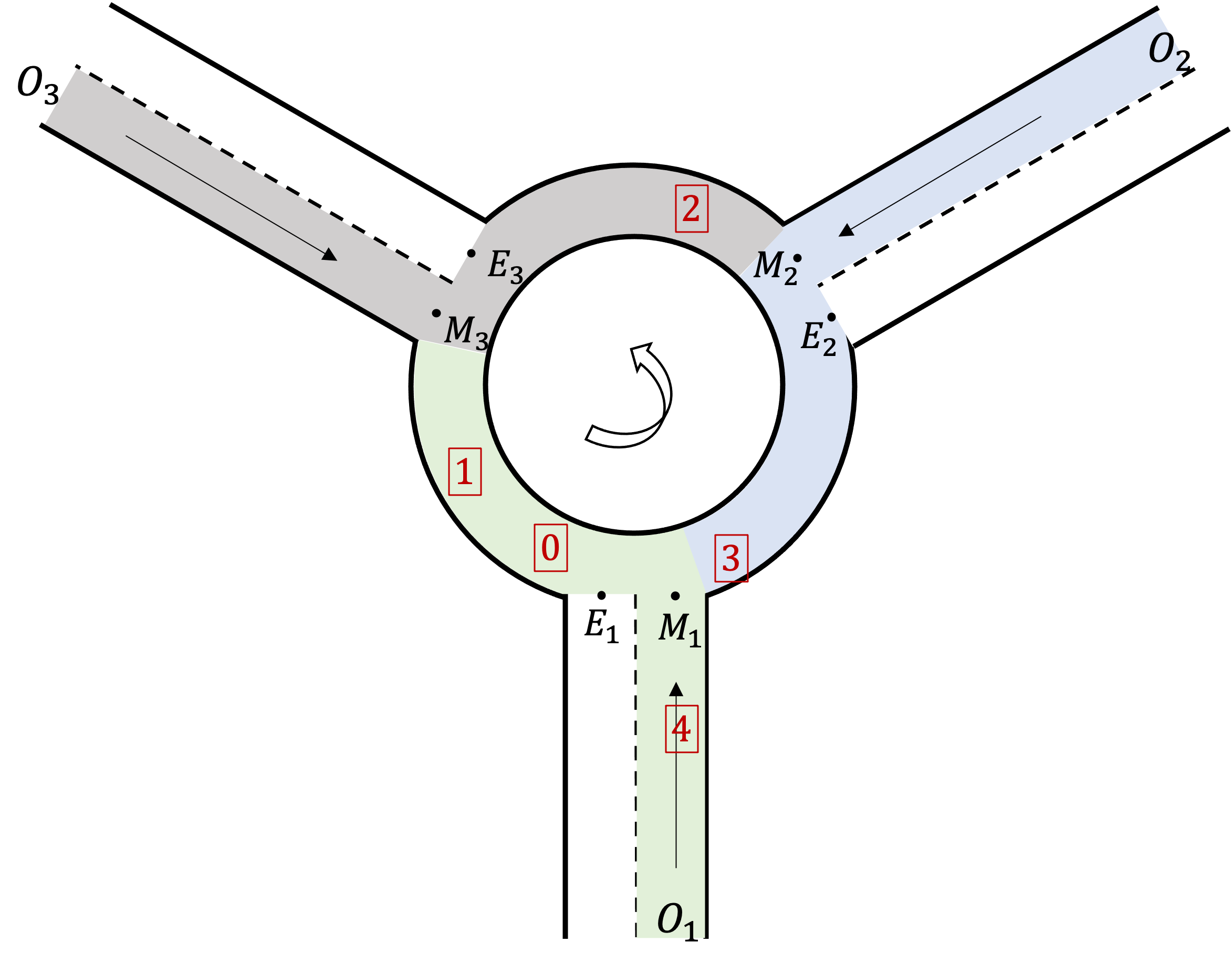}
    \caption{A roundabout with 3 entries }
    \label{fig:roundabout}
\end{figure}
The objective is to develop controllers for Connected and Automated Vehicles (CAVs) traversing a single-lane roundabout so as to simultaneously determine the optimal merging sequence and the associated optimal motion control that jointly minimize travel time and energy consumption, while guaranteeing speed-dependent safety and satisfying velocity and acceleration constraints. 

As in \cite{chen2025optimal}, we consider a single-lane three-entry roundabout consisting of three \emph{Control Zones} (CZs), each centered at a merging point, as shown in Fig.~\ref{fig:roundabout}. CAVs enter the roundabout randomly from different entry roads and move counterclockwise through multiple interconnected CZs before exiting.
The longitudinal dynamics of each CAV $i$ are modeled as
\begin{equation}
\left[
\begin{array}{c}
\dot{x}_{i}(t)\\
\dot{v}_{i}(t)
\end{array}
\right]
=
\left[
\begin{array}{c}
v_{i}(t)\\
u_{i}(t)
\end{array}
\right],
\label{VehicleDynamics}
\end{equation}
where $x_i(t)$ denotes the distance to the origin of the current road segment, $v_i(t)$ is the velocity, and $u_i(t)$ is the control input (acceleration). Let $\bm{x}_i(t)=[x_i(t), v_i(t)]^T$ denotes the state vector.

We impose four classes of safety constraints in the roundabout coordination problem. 
For state-dependent constraints that are satisfied at initialization, safety is enforced through the CBF condition \eqref{eq:cbf}. 
For completeness, we present both the original constraint formulation and the corresponding CBF transformation.

\emph{(i) Vehicle Limitation Constraints:}
\begin{equation}\setlength{\abovedisplayskip}{1pt}\setlength{\belowdisplayskip}{1pt}\label{eq:roundabout-control limit}
u_{i,\min} \le u_i(t) \le u_{i,\max}, 
\end{equation}
\begin{equation}\setlength{\abovedisplayskip}{1pt}\setlength{\belowdisplayskip}{1pt}\label{eq:roundabout-velocity limit}
v_{\min} \le v_i(t) \le v_{\max},
\end{equation}
where $v_{\max}>0$ and $v_{\min}\ge 0$ denote allowable speed bounds, and $u_{i,\min}<0$, $u_{i,\max}>0$ are acceleration limits. 
These bounds are enforced through the CBF constraints:
\begin{equation}\setlength{\abovedisplayskip}{1pt}\setlength{\belowdisplayskip}{1pt}\label{eqn: CBF_velocity_1}
    -u_{i}(t) + \alpha_1(b_1(\bm{x}_{i}(t)))\geq 0
\end{equation} 
\begin{equation}\setlength{\abovedisplayskip}{1pt}\setlength{\belowdisplayskip}{1pt}\label{eqn: CBF_velocity_2}
    u_{i}(t) + \alpha_2(b_2(\bm{x}_{i}(t)))\geq 0
\end{equation} 
where $b_1(\bm{x}_{i}(t))=v_{max} - v_{i}(t)$, $b_2(\bm{x}_{i}(t)) = v_{i}(t) - v_{min}$.

\emph{(ii) Lateral Safety Constraint:} When traversing the roundabout, the centrifugal force acting on the vehicle must remain sufficiently small to prevent rollover. 
This requirement is expressed as
\begin{equation}\setlength{\abovedisplayskip}{2pt}\setlength{\belowdisplayskip}{2pt}
\kappa_i(d_i(t))\cdot v_i^2(t)\cdot h \leq w_h\cdot g \label{equ: lateral safe}
\end{equation}
where $h$ is the vehicle height, $w_h$ is the half width of the vehicle (assumed identical for all CAVs), and $g$ is the gravitational constant. 
The corresponding CBF condition becomes
\begin{equation}\setlength{\abovedisplayskip}{1pt}\setlength{\belowdisplayskip}{1pt}\label{eqn: CBF_lateral}
    -2\kappa_i(d_{i}(t)) h v_{i}(t) u_{i}(t)  + \alpha_3(b_3(\bm{x}_{i}(t))) \geq 0
\end{equation}
where $b_3(\bm{x}_{i}(t))=w_h\cdot g -\kappa_i(d_{i}(t))\cdot v_{i}(t)^2\cdot h $.

\emph{(iii) Rear-end Safety Constraint:}To avoid rear-end collisions, each vehicle must maintain a speed-dependent distance from its immediately preceding vehicle $i_p$:
\begin{equation}\label{eq:roundabout-rearend constraint}
x_{i_p}(t) - x_i(t) \ge \varphi v_i(t) + \delta,
\end{equation}
where $\varphi$ represents a reaction time parameter and $\delta$ accounts for vehicle length. 
The corresponding CBF constraint is
\begin{equation}\setlength{\abovedisplayskip}{1pt}\setlength{\belowdisplayskip}{1pt}\label{eqn: CBF_rearend}
    v_{i_p}(t) - v_{i}(t) - \varphi u_{i}(t) + \alpha_4(b_4(\bm{x}_{i}(t))) \geq 0
\end{equation}
where $b_4(\bm{x}_{i}(t))=x_{i_p}(t) - x_i(t)-\varphi v_{i}(t) -\delta $.

\emph{(iv) Safe-merging Constraint:}
At a merging point $M_k$, vehicle $i$ must maintain a sufficient separation from its conflicting vehicle $i_m$ 
(the optimal choice of $i_m$ is presented in \cite{chen2025optimal}).
Let $z_{i,i_m}(t):=x_{i_m}(t)-x_i(t)$ denote their relative distance along the corresponding approaches. 
The merging safety requirement is expressed as
\begin{equation}\label{eq:roundabout-safe_merging constraint}
z_{i,i_m}(t)\ge \frac{\varphi\, x_{i_m}(t)}{L_{i_m}}\, v_{i}(t)+\delta,
\qquad \forall t\in [t_i^{k,0},\, t_{i_m}^k],
\end{equation}
where $t_i^{k,0}$ is the time CAV $i$ enters the segment leading to $M_k$ and $L_{i_m}$ is the segment length. 
The corresponding CBF constraint becomes
{\small\begin{align}\setlength{\abovedisplayskip}{1pt}\setlength{\belowdisplayskip}{1pt}\label{eq:cbf_merging}
    &v_{i_m}(t)-v_i(t)-\frac{\varphi}{L_{i_m}}x_{i_m}(t)u_i(t) -\frac{\varphi}{L_{i_m}}v_{i_m}(t)v_i(t)\nonumber \\ 
    & + \alpha_5(b_5(\bm x_{i}(t))) \geq 0, ~~\forall t\in [t_i^{k,0}, t_{i_m}^k]
\end{align}}
where 
{\small\begin{align}\setlength{\abovedisplayskip}{1pt}\setlength{\belowdisplayskip}{1pt}\label{eq:b_5 merging}
    b_5(\bm{x}_{i}(t))=(L_i-x_{i}(t)) - &(L_{i_m} - x_{i_m}(t))\nonumber\\&-\frac{\varphi}{L_{i_m}}x_{i_m}(t)\cdot v_{i}(t) -\delta
\end{align}}

Replacing the original constraints with their CBF counterparts preserves the safety requirements and guarantees forward invariance of the safe set while producing constraints that are affine in the control input. 
This structure enables efficient real time implementation through quadratic programming (QP) 
(details in \cite{chen2025optimal}).
The degree of conservativeness depends on the choice of the class-$\mathcal{K}$ functions $\alpha_i$, $i\in\{1,\ldots,5\}$.

A limitation of standard CBF formulations is that they require the safety condition to hold at initialization. 
In practice, due to stochastic arrivals and upstream decisions, the safe-merging condition \eqref{eq:roundabout-safe_merging constraint} may be violated when a vehicle enters a new control zone within the roundabout. 
In this case, the standard CBF constraint is no longer applicable. 
To address this situation, we replace the merging constraint with an ExT-CBF constraint, which enforces recovery within a prescribed time horizon
to restore the safety before the conflicting vehicle reaches the merging point.

Specifically, if $b_5(\bm{x}(t_0))<0$ at the current decision time $t_0$, an exact-time recovery mechanism is activated. 
The recovery horizon $T_r$ is chosen as the nominal time required for vehicle $i_m$ to reach the merging point under its current motion, thereby providing a physically meaningful time window for safety restoration.

We adopt a quadratic recovery function
\begin{equation}\setlength{\abovedisplayskip}{1pt}\setlength{\belowdisplayskip}{1pt}\label{roundabout:gamma}
\gamma_i(t)= b_5(\bm{x}(t_0))\!\left(1-a\tau-b\tau^2\right),
\qquad 
\tau=\frac{t-t_0}{T_r},
\end{equation}
where the shape parameters $a,b$ are decision variables optimized through \textbf{Problem 1} to \emph{balance travel time and energy consumption.} 
Let 
\begin{align}\label{eq:ext_b5}
b^{\small ExT}_5(t) = &L_f b_5(\bm{x}(t))+L_g b_5(\bm{x}(t))u_i(t)-\dot{\gamma}_i(t)\nonumber\\
& +\alpha_5\!\left(b_5(\bm{x}(t))-\gamma_i(t)\right) \nonumber\\
=&\, v_{i_m}(t)-v_i(t)
-\frac{\varphi}{L_{i_m}}x_{i_m}(t)u_i(t)
-\frac{\varphi}{L_{i_m}}v_{i_m}(t)v_i(t) \nonumber\\
&-\dot{\gamma}_i(t)
+\alpha_5\!\left(b_5(\bm{x}(t))-\gamma_i(t)\right)  
\end{align}

Over the interval $[t_0,t_0+T_r]$, the merging constraint is replaced by the ExT-CBF tracking condition
\begin{align}
\label{roundabout:ext-cbf safe-merging constraint}
b^{\small ExT}_5(t)
= \delta_r(t),
\quad t\in[t_0,t_0+T_r],
\end{align}
where $\delta_r(t)\ge 0$ is a slack variable penalized in the objective to promote exact barrier tracking. 
When $\delta_r(t)=0$, the barrier follows the prescribed trajectory $\gamma_i(t)$ and satisfies $b_i(t_0+T_r)=0$ by construction. 
For $t\ge t_0+T_r$, a standard CBF condition is imposed to guarantee forward invariance thereafter. 
This formulation embeds the desired recovery time directly into the barrier dynamics while preserving feasibility under input constraints.

Given any merging sequence, the motion control of a CAV is determined by solving a QP exploiting the affine dependence on $u$ in \eqref{eq:cbf}. 
This is denoted as \textbf{Problem 2}:
\begin{subequations}\label{eq:problem2}
{\small
\begin{align}
\min_{u_i(t),\,\delta_r(t)} 
& J_i(t) =
\frac{\frac{1}{2}u_{i}^{2}(t)}{\max\{u_{i,\max}^2,u_{i,\min}^2\}}
+\lambda_1 \frac{(v_i(t)-v_d)^2}{(v_{\max}-v_{\min})^2} \notag\\
&\hspace{3.5em}
+\lambda_2 \frac{\kappa_i(d_i(t))v_i(t)^2}{\kappa_{\max}v_{\max}^2}
+\lambda_3 \delta_r(t)^2
\label{eq:problem2_objective}\\
\text{s.t.}\quad
& \eqref{VehicleDynamics}, \eqref{eq:roundabout-control limit}, \eqref{eqn: CBF_velocity_1}, \eqref{eqn: CBF_velocity_2}, \eqref{eqn: CBF_lateral}, \eqref{eqn: CBF_rearend}, \notag\\
& \begin{cases}
\text{\eqref{roundabout:ext-cbf safe-merging constraint}},
& \text{if } b_i(t_0) < 0, \\
\eqref{eq:cbf_merging},
& \text{otherwise}.
\end{cases}
\end{align}}
\end{subequations}
where $\lambda_1, \lambda_2\ge 0$ balance energy consumption, speed regulation, and comfort, while $\lambda_3>0$ penalizes deviations from exact barrier tracking. 
Here $v_d$ denotes the desired speed and $\kappa_{max}$ is the maximum curvature of the roundabout.

The detailed optimal sequencing procedure, including the discretization and MPC implementation used for sequence evaluation and selection, is detailed in \cite{chen2025optimal}.

\section{Simulation Results}

We evaluate the proposed Exact-time Recovery CBF through two case studies.

The first case considers a simple ACC scenario to demonstrate the core mechanism of the ExT-CBF method. 
In particular, it illustrates the Exact Barrier Tracking capability and shows how the barrier trajectory can be optimally shaped while enforcing recovery at a prescribed time.

The second case examines 
the roundabout coordination problem of Section \ref{Sec:Optimal Control for Roundabout} studied in \cite{chen2025optimal}. 
Here, the Exact-time Recovery CBF replaces the CLBF-based finite-time mechanism to enforce safe merging constraints. 
This example demonstrates that the proposed framework extends to complex traffic networks, improving performance and reducing infeasibility under tight safety and input constraints.

\subsection{Adaptive Cruise Control}

We first consider a classical adaptive cruise control (ACC) scenario. 
The longitudinal dynamics are modeled as a double integrator $
\dot{x} = v, 
\dot{v} = u
$,
where $x$ denotes the relative position, $v$ is the ego vehicle speed, and $u$ is the longitudinal acceleration input.

The safety constraint is defined through the barrier function
$b(x) = z - \phi v$,
where $z$ is the inter-vehicle distance and $\phi>0$ denotes the time headway parameter. 
The corresponding safe set is $\{x \mid b(x) \ge 0\}$.

With the initial conditions
$
z(t_0)=20, 
v(t_0)=20, 
v_p(t_0)=15,
$
the safety constraint is violated at $t=t_0$, resulting in $b(t_0) = -16$.
The prescribed recovery horizon is set to $T=15$, meaning that the controller is required to steer the system back to the safe set within $15$ time units.

Three different controllers are adopted for comparison:

\textbf{(a) Exact-time recovery CBF.}  
We parameterize the recovery trajectory $\gamma(t;\theta)$ using a quadratic recovery function of the form
\[
\gamma(t) = a \left(\frac{t}{T}\right)^2 + b \left(\frac{t}{T}\right) + c.
\]
When the recovery constraint is active, the barrier satisfies
\[
b(x(t)) = \gamma(t), 
\quad t \in [t_0,t_0+T],
\]
so recovery occurs exactly at $T$ by construction. 
The parameters $\theta$ are obtained by solving \textbf{Problem 1} under two objectives:  
(i) minimizing control effort, and  
(ii) minimizing terminal speed deviation.
\medskip

\textbf{(b) Control Lyapunov Barrier Function (CLBF)}  \cite{xiao2022hocbf} with $p=0.766$, $q=\frac{1}{5}$. 
The CLBF enforces $\dot b(x) \geq p(-b(x))^{q}$, yielding the nominal convergence time
\[
t_{\mathrm{conv}}
= \frac{b(0)^{1-q}}{p(1-q)}
\approx T.
\]
This guarantees recovery no later than $T$, but not exactly at $T$. 
The fixed polynomial rate may induce conservative control action.

\medskip

\textbf{(c) FxT-CBF}  \cite{garg2021robust}. 
The FxT-CBF \ combines a CBF safety constraint with a fixed-time Lyapunov structure
\[
\dot V(x) 
\le 
- c_1 V^{\gamma_1} 
- c_2 V^{\gamma_2},
\qquad 
\gamma_1 = 1 + \tfrac{1}{\mu}, 
\quad 
\gamma_2 = 1 - \tfrac{1}{\mu}.
\]
This guarantees convergence within a time bound determined solely by the design parameters $(c_1,c_2,\mu)$ and independent of initial conditions inside a prescribed domain. 
Following the implementation in \cite{garg2021robust}, we select
\begin{equation}\label{eq:FxT-CBF params}
    c_1 = c_2 = \frac{\mu \pi}{2T}, 
\qquad 
\mu = 2
\end{equation}
so that the resulting time bound matches the desired $T$.

Unlike the exact-time recovery CBF, the FxT-CBF provides a guaranteed upper bound on the convergence time, but does not enforce recovery precisely at $T$. As a result, the barrier evolution is not explicitly prescribed and may reach the safe set strictly before the deadline.

\begin{figure}
    \centering
    \includegraphics[width=0.9\columnwidth]{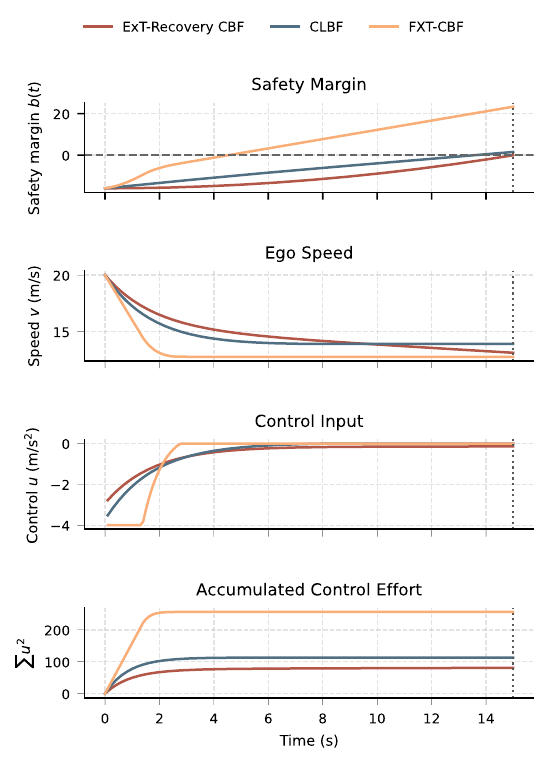}
    \caption{ACC comparison with ExT-CBF optimized for \emph{accumulated control effort}.}
    \label{fig:acc_opt_control}
\end{figure}

\begin{figure}
    \centering
    \includegraphics[width=0.99\columnwidth]{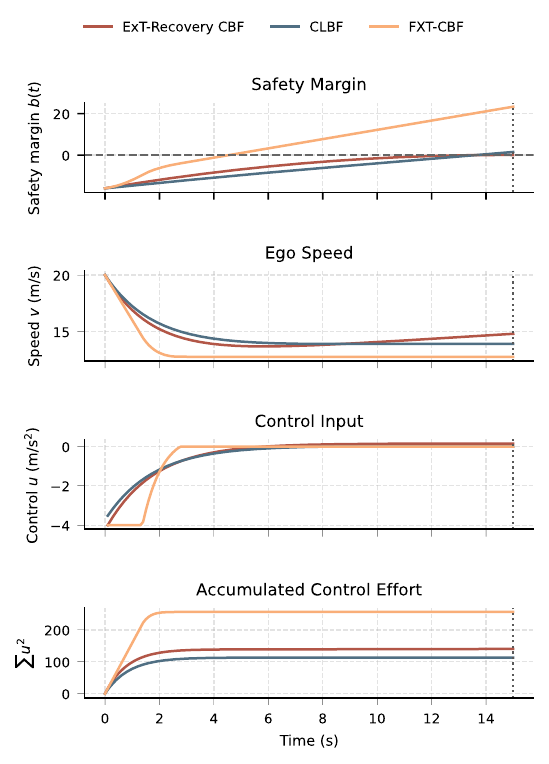}
    \caption{ ACC comparison with ExT-CBF optimized for  \emph{terminal speed deviation}.}
    \label{fig:acc_opt_speed}
\end{figure}

The comparison results are shown in Figs. \ref{fig:acc_opt_control} and \ref{fig:acc_opt_speed}.
In both cases, the Exact-time recovery CBF achieves recovery exactly at $t=T=15$, as clearly shown in the safety margin subplots where $b(t)$ reaches zero precisely at the prescribed deadline. 
In contrast, both CLBF and FxT-CBF exhibit conservative behavior: the barrier trajectory crosses zero strictly before $T$, which exceeds the requirement of convergence at $T$ and leads to unnecessarily aggressive early control corrections.

\noindent

Moreover, while enforcing this exact recovery constraint, the proposed method simultaneously optimizes the prescribed performance objective. 
When minimizing accumulated control effort (Fig. \ref{fig:acc_opt_control}), the ExT-CBF controller yields a substantially lower $\sum u^2$, approximately $25\%$ lower than CLBF and about $65\%$ lower than FxT-CBF. 
When optimizing terminal speed deviation (Fig. \ref{fig:acc_opt_speed}), an interesting transient behavior emerges: the controller initially decelerates to rapidly mitigate the most of the safety violation, then smoothly accelerates to meet the terminal objective at $T$. 
This illustrates that the barrier trajectory is not merely a feasibility constraint, but an actively shaping design variable that balances safety recovery and performance.

\subsection{Roundabout Control}
The setting and control framework follows Section~\ref{Sec:Optimal Control for Roundabout}. We adopt the same optimal sequencing strategy as in~\cite{chen2025optimal}, where each candidate sequence is evaluated within a receding-horizon framework consisting of a standard QP with CBF constraints to achieve optimal performance while enforcing safety requirements. The only difference lies in how the controller responds when the merging constraint is violated.

We compare four recovery mechanisms for handling this situation.
(a) \emph{CBF only:} Any initial conflict renders the problem infeasible; therefore, the controller applies the minimum admissible control to leave the unsafe state as quickly as possible.
(b) \emph{CLBF:} The decay parameters are determined through an auxiliary optimization that maximizes the feasible control region while ensuring safety convergence before the actual merging event, as proposed in~\cite{chen2025optimal}.
(c) \emph{FxT-CBF}~\cite{garg2021robust}: The parameters are selected using the same framework as in~\eqref{eq:FxT-CBF params} to guarantee convergence within a fixed-time bound.
(d) \emph{ExT-CBF:} The control input is obtained by solving \textbf{Problem~2}, where the barrier evolution follows a quadratic recovery trajectory defined in~\eqref{roundabout:gamma}.

System performance is evaluated using two categories of metrics: \emph{efficiency} (energy consumption, travel time, and passenger discomfort) and \emph{safety} (potential risks both within the roundabout and along incoming approaches). 
The ``infeasible count'' is the average number of time steps per CAV for which no conflict-free merging sequence exists, and the ``hard deceleration count'' is the average number of time steps per CAV for which the maximum allowable deceleration is applied to resolve conflicts.

The baseline parameter settings are
$
L_k^0 = L_k^1 = 60\,\text{m}, \forall k \in \{1,2,3\}, 
\delta = 0\,\text{m}, 
\varphi = 1.8\,\text{s},
$,
$
v_{\max} = 20\,\text{m/s}, \quad v_{\min} = 0\,\text{m/s}, \quad
u_{i,\max} = 4\,\text{m/s}^2, \quad
u_{i,\min} = -4\,\text{m/s}^2,
$
for all vehicles \(i\).

\begin{figure}
    \centering
    \includegraphics[width=0.99\columnwidth]{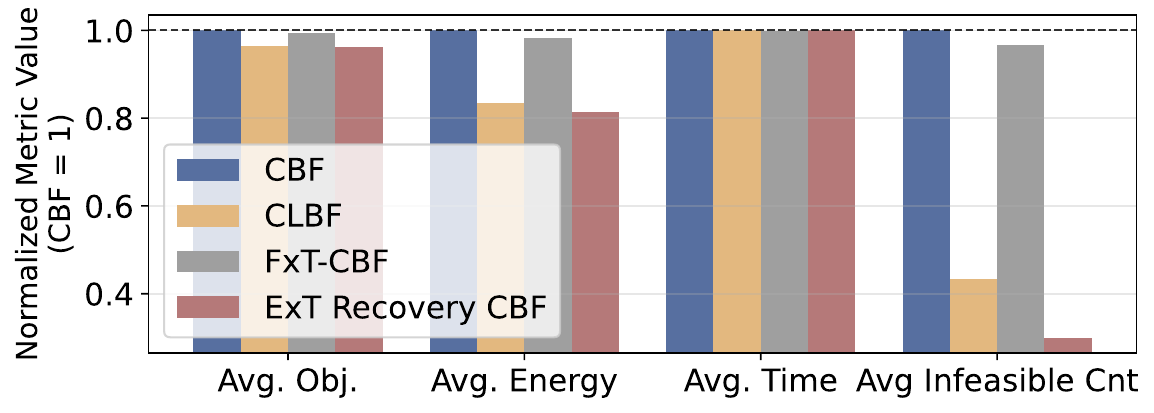}
    \caption{Relative Performance Across Methods for Balanced Roundabout Traffic}
    \label{fig:roundabout_metric_comparison}
\end{figure}

Figure~\ref{fig:roundabout_metric_comparison} presents the normalized performance comparison under balanced traffic, where each metric is scaled with respect to the CBF-only baseline. Although recovery is triggered only when stochastic vehicle arrivals or transitions between control zones violate the merging constraint,
clear performance differences emerge during these events. The ExT-CBF controller achieves the lowest normalized objective and energy among all methods. In particular, the reduction in average energy reflects decreased conservativeness during recovery. By prescribing the barrier trajectory to reach the safe set exactly at the desired time, the controller avoids unnecessarily aggressive early corrections that arise in CLBF and FxT-CBF due to their fixed convergence structures. Moreover, the infeasible count is further reduced under ExT-CBF, indicating improved compatibility with input constraints. This improvement is attributable to the greater flexibility of the parameterized recovery profile, which enlarges the feasible region compared to the fixed polynomial or dual-power convergence laws.

Extending to all traffic scenarios, the quantitative results in Table~\ref{tab:perf_all} confirm the same trend. Across balanced, unbalanced, and heavy traffic, the Exact-time Recovery CBF consistently yields the lowest overall objective and reduced energy consumption, while also lowering infeasibility relative to OCBF and FxT-CBF. Even though absolute differences remain moderate due to the sporadic activation of recovery constraints, the proposed framework systematically reduces conservativeness and improves feasibility without sacrificing safety performance.

\begin{table*}
\centering
\caption{Performance comparison under balanced, unbalanced, and heavy traffic scenarios.}
\label{tab:perf_all}
\begin{threeparttable}
\setlength{\tabcolsep}{3pt}
\renewcommand{\arraystretch}{1.05}
\begin{tabular}{|l|cccc|cccc|cccc|}
\hline
 & \multicolumn{4}{c|}{Balanced} & \multicolumn{4}{c|}{Unbalanced} & \multicolumn{4}{c|}{Heavy} \\
\cline{2-13}
Metric
& CBF & CLBF & FxT-CBF & \textbf{ExT-CBF}
& CBF & CLBF & FxT-CBF & \textbf{ExT-CBF}
& CBF & CLBF & FxT-CBF & \textbf{ExT-CBF} \\
\hline
Avg. Obj. \eqref{eq:problem2_objective}
& 0.990 & 0.954 & 0.984 & \textbf{0.952}
& 1.091 & 0.976 & 1.034 & \textbf{0.974}
& 1.045 & 0.989 & 1.010 & \textbf{0.985} \\

Avg. Energy
& 2.293 & 1.915 & 2.252 & \textbf{1.866}
& 3.210 & 2.081 & 2.731 & \textbf{1.965}
& 2.865 & 2.240 & 2.492 & \textbf{2.179} \\

Avg. Time
& 16.468 & 16.466 & \textbf{16.459} & 16.468
& 16.502 & 16.424 & \textbf{16.415} & 16.462
& 16.174 & 16.194 & \textbf{16.180} & 16.195 \\

Avg. Discomfort
& 48.621 & \textbf{48.549} & 48.579 & 48.692
& 48.031 & \textbf{48.029} & 48.103 & 48.107
& 46.824 & \textbf{46.719} & 46.729 & 46.770 \\
\hline
Avg. Hard Deceleration Cnt
& 0.536 & 0.232 & 0.518 & \textbf{0.161}
& 1.255 & 0.418 & 1.000 & \textbf{0.309}
& 0.963 &  \textbf{0.375} & 0.613 & \textbf{0.375} \\

Avg Infeasible Cnt
& 0.536 & 0.232 & 0.518 & \textbf{0.161}
& 1.255 & 0.418 & 1.000 & \textbf{0.309}
& 0.725 &  \textbf{0.213} & 0.450 & \textbf{0.213} \\
\hline
\end{tabular}

\end{threeparttable}
\vspace{-1em}
\end{table*}


\section{Conclusion and Future Work}

This paper introduced an exact-time safety recovery framework for control-affine systems with unsafe initial conditions using time-varying control barrier functions and optimal barrier tracking. By prescribing and optimizing a recovery trajectory in barrier space, the proposed ExT-CBF framework guarantees recovery to the safe set at a desired time when feasible under input constraints, while reducing the conservativeness associated with conventional finite-time recovery laws. The roundabout coordination results showed that this additional flexibility improves performance and feasibility.
Future work will focus on extending the framework to higher-relative-degree constraints, incorporating robustness to uncertainty and disturbances.

\bibliographystyle{IEEEtran}
\bibliography{ref}

\begin{thebibliography}{10}
\providecommand{\url}[1]{#1}
\csname url@samestyle\endcsname
\providecommand{\newblock}{\relax}
\providecommand{\bibinfo}[2]{#2}
\providecommand{\BIBentrySTDinterwordspacing}{\spaceskip=0pt\relax}
\providecommand{\BIBentryALTinterwordstretchfactor}{4}
\providecommand{\BIBentryALTinterwordspacing}{\spaceskip=\fontdimen2\font plus
\BIBentryALTinterwordstretchfactor\fontdimen3\font minus \fontdimen4\font\relax}
\providecommand{\BIBforeignlanguage}[2]{{%
\expandafter\ifx\csname l@#1\endcsname\relax
\typeout{** WARNING: IEEEtran.bst: No hyphenation pattern has been}%
\typeout{** loaded for the language `#1'. Using the pattern for}%
\typeout{** the default language instead.}%
\else
\language=\csname l@#1\endcsname
\fi
#2}}
\providecommand{\BIBdecl}{\relax}
\BIBdecl

\bibitem{liu2025serverless}
S.~Liu and C.~A. Shue, ``Functional control: Leveraging function-as-a-service platforms for software-defined networking controllers,'' in \emph{Proceedings of the Symposium on Theory, Algorithmic Foundations, and Protocol Design for Mobile Networks and Mobile Computing}, ser. MobiHoc '25.\hskip 1em plus 0.5em minus 0.4em\relax New York, NY, USA: ACM, 2025, p. 141–150.

\bibitem{rios2017survey}
J.~Rios-Torres and A.~A. Malikopoulos, ``A survey on the coordination of connected and automated vehicles at intersections and merging at highway on-ramps,'' \emph{IEEE Transactions on Intelligent Transportation Systems}, vol.~18, no.~5, pp. 1066--1077, 2017.

\bibitem{rios2017merging}
------, ``Automated and cooperative vehicle merging at highway on-ramps,'' \emph{IEEE Transactions on Intelligent Transportation Systems}, vol.~18, no.~4, pp. 780--789, 2017.

\bibitem{campi2023roundabouts}
E.~Campi, G.~Mastinu, G.~Previati, L.~Studer, and L.~Uccello, ``Roundabouts: Traffic simulations of connected and automated vehicles—a state of the art,'' \emph{IEEE Trans. on Intelligent Transp. Systems}, pp. 1--21, 2023.

\bibitem{nagumo1942lage}
M.~Nagumo, ``\"uber die lage der integralkurven gew{\"o}hnlicher differentialgleichungen,'' \emph{Proceedings of the Physico-Mathematical Society of Japan. 3rd Series}, vol.~24, pp. 551--559, 1942.

\bibitem{Ames2017TAC}
A.~D. Ames, X.~Xu, J.~W. Grizzle, and P.~Tabuada, ``Control barrier function based quadratic programs for safety critical systems,'' \emph{IEEE Transactions on Automatic Control}, vol.~62, no.~8, pp. 3861--3876, 2016.

\bibitem{glotfelter2017nonsmooth}
P.~Glotfelter, J.~Cort{\'e}s, and M.~Egerstedt, ``Nonsmooth barrier functions with applications to multi-robot systems,'' \emph{IEEE Control Systems Letters}, vol.~1, no.~2, pp. 310--315, 2017.

\bibitem{nguyen2016exponential}
Q.~Nguyen and K.~Sreenath, ``Exponential control barrier functions for enforcing high relative-degree safety-critical constraints,'' in \emph{2016 American Control Conference (ACC)}.\hskip 1em plus 0.5em minus 0.4em\relax IEEE, 2016, pp. 322--328.

\bibitem{xiao2022hocbf}
W.~Xiao and C.~Belta, ``High-order control barrier functions,'' \emph{IEEE Transactions on Automatic Control}, vol.~67, no.~7, pp. 3655--3662, 2022.

\bibitem{milanes2010automated}
V.~Milan{\'e}s, J.~Godoy, J.~Villagr{\'a}, and J.~P{\'e}rez, ``Automated on-ramp merging system for congested traffic situations,'' \emph{IEEE Trans. on Intelligent Transp. Systems}, vol.~12, no.~2, pp. 500--508, 2010.

\bibitem{li2025robust}
A.~Li, A.~S.~C. Armijos, and C.~G. Cassandras, ``Robust optimal lane-changing control for connected autonomous vehicles in mixed traffic,'' \emph{Automatica}, vol. 174, p. 112169, 2025.

\bibitem{xu2022decentralized}
K.~Xu, C.~G. Cassandras, and W.~Xiao, ``Decentralized time and energy-optimal control of connected and automated vehicles in a roundabout with safety and comfort guarantees,'' \emph{IEEE Trans. on Intelligent Transp. Systems}, vol.~24, no.~1, pp. 657--672, 2022.

\bibitem{chen2021backup}
Y.~Chen, M.~Jankovic, M.~Santillo, and A.~D. Ames, ``Backup control barrier functions: Formulation and comparative study,'' in \emph{2021 60th IEEE Conference on Decision and Control (CDC)}.\hskip 1em plus 0.5em minus 0.4em\relax IEEE, 2021, pp. 6835--6841.

\bibitem{janwani2024learning}
N.~C. Janwani, E.~Da{\c{s}}, T.~Touma, S.~X. Wei, T.~G. Molnar, and J.~W. Burdick, ``A learning-based framework for safe human-robot collaboration with multiple backup control barrier functions,'' in \emph{2024 IEEE International Conference on Robotics and Automation (ICRA)}.\hskip 1em plus 0.5em minus 0.4em\relax IEEE, 2024, pp. 11\,676--11\,682.

\bibitem{cosner2021measurement}
R.~K. Cosner, A.~W. Singletary, A.~J. Taylor, T.~G. Molnar, K.~L. Bouman, and A.~D. Ames, ``Measurement-robust control barrier functions: Certainty in safety with uncertainty in state,'' in \emph{2021 IEEE/RSJ International Conference on Intelligent Robots and Systems (IROS)}.\hskip 1em plus 0.5em minus 0.4em\relax IEEE, 2021, pp. 6286--6291.

\bibitem{garg2021robust}
K.~Garg and D.~Panagou, ``Robust control barrier and control lyapunov functions with fixed-time convergence guarantees,'' in \emph{2021 American Control Conference (ACC)}.\hskip 1em plus 0.5em minus 0.4em\relax IEEE, 2021, pp. 2292--2297.

\bibitem{zhang2024prescribed}
S.~Zhang, D.-H. Zhai, Y.~Xiong, and Y.~Xia, ``Prescribed-time safety control for unknown systems and its application to robotic manipulator,'' \emph{IEEE Transactions on Automation Science and Engineering}, vol.~22, pp. 9923--9933, 2024.

\bibitem{huang2024learning}
P.~Huang, F.~Yao, Q.~Lu, W.~Pan, and L.~Wang, ``Learning-based prescribed-time safety for control of unknown systems with control barrier functions,'' \emph{IEEE Control Systems Letters}, vol.~8, pp. 2439--2444, 2024.

\bibitem{xiao2021clbf}
W.~Xiao, C.~A. Belta, and C.~G. Cassandras, ``High order control lyapunov-barrier functions for temporal logic specifications,'' in \emph{2021 American Control Conference (ACC)}.\hskip 1em plus 0.5em minus 0.4em\relax IEEE, 2021, pp. 4886--4891.

\bibitem{chen2025optimal}
Y.~Chen and C.~G. Cassandras, ``Optimal sequencing and motion control in a roundabout with safety and comfort guarantees,'' \emph{IEEE Transactions on Intelligent Transportation Systems}, vol.~26, no.~11, pp. 19\,148--19\,162, 2025.

\bibitem{xiao2023safe}
W.~Xiao, C.~G. Cassandras, and C.~Belta, \emph{Safe autonomy with control barrier functions: Theory and applications}.\hskip 1em plus 0.5em minus 0.4em\relax Springer, 2023.

\bibitem{li2026finitetimeconvergentcontrolbarrier}
\BIBentryALTinterwordspacing
A.~Li, Y.~Chen, C.~G. Cassandras, and W.~Xiao, ``Finite-time convergent control barrier functions with feasibility guarantees,'' 2026. [Online]. Available: \url{https://arxiv.org/abs/2603.22445}
\BIBentrySTDinterwordspacing

\bibitem{gadginmath2026constrictingtubesprescribedtimesafe}
\BIBentryALTinterwordspacing
D.~Gadginmath, A.~Allibhoy, and F.~Pasqualetti, ``Constricting tubes for prescribed-time safe control,'' 2026. [Online]. Available: \url{https://arxiv.org/abs/2603.17003}
\BIBentrySTDinterwordspacing

\end{thebibliography}
\end{document}